\documentclass[final,3p,times,twocolumn]{elsarticle}
\usepackage{graphicx}
\usepackage{array}
\usepackage{algorithm2e}

\usepackage{amssymb}
\usepackage{amsmath}
\usepackage{bm}
\usepackage{amsthm}
\usepackage{dirtree}
\usepackage{lmodern}

\usepackage{listings}
\newcommand{\belowtitleskip}{2pt}%\smallskipamount}

\usepackage{xcolor}
\definecolor{gray}{gray}{0.97}
\colorlet{commentcolour}{green!50!black}
\colorlet{stringcolour}{red!60!black}
\colorlet{keywordcolour}{magenta!90!black}
\colorlet{exceptioncolour}{yellow!50!red}
\colorlet{commandcolour}{blue!60!black}
\colorlet{promptcolour}{green!50!black}

\lstdefinestyle{pythonstyle}{
keepspaces=true,
language=python,
showtabs=true,
tab=,
tabsize=2,
basicstyle=\ttfamily\footnotesize,%\setstretch{.5},
stringstyle=\color{stringcolour},
showstringspaces=false,
alsoletter={1234567890},
otherkeywords={\ , \}, \{, \%, \&, \|},
keywordstyle=\color{keywordcolour}\bfseries,%
emph={[2]True, False, None},
emphstyle=[2]\color{keywordcolour},
emph={[3]object,type,isinstance,copy,deepcopy,zip,enumerate,
reversed,list,len,dict,tuple,xrange,append,execfile,real,imag,
reduce,str,repr, vars},
emphstyle=[3]\color{commandcolour},
emph={Exception,NameError,IndexError,SyntaxError,TypeError,ValueError,OverflowError,ZeroDivisionError},
emphstyle=\color{exceptioncolour}\bfseries,
morestring=[s]{"""}{"""},
morestring=[s]{'''}{'''},
commentstyle=\color{commentcolour}\slshape,
emph={[4]1, 2, 3, 4, 5, 6, 7, 8, 9, 0, 001, ode, fsolve, sqrt, exp, sin, cos, arccos, pi, array, norm, dot, arange, , isscalar, max, sum, flatten, shape, reshape, find, any, all, abs, plot, linspace, legend, quad, polyval,polyfit, hstack, concatenate,vstack,column_stack,empty,zeros,ones,rand,vander,
grid,pcolor,eig,eigs,eigvals,svd,qr,tan,det,logspace,roll,min,
mean,cumsum,cumprod,diff,vectorize,lstsq,cla,eye,xlabel,ylabel,
},
emphstyle=[4]\color{commandcolour},
emph={[5]and,break,class,continue,def,yield,del,elif ,else,
except,exec,finally,for,global,if,in,
lambda,not,or,pass,print,raise,return,try,while,assert},
emphstyle=[5]\color{black}\bf,
literate=*%
{>>>}{{\textcolor{promptcolour}{>>>}}}{1}%
,%
breaklines=true,
breakatwhitespace= true,
xleftmargin=0ex,
xrightmargin=0ex,
aboveskip=1ex,
belowskip=1ex,
frame=trbl, %trbl
numbers=none,
rulecolor=\color{black!40},
backgroundcolor=\color{gray}
}

\lstdefinestyle{inlinestyle}{
language=python,
basicstyle=\ttfamily\footnotesize,%\setstretch{.5},
breaklines=true,
breakatwhitespace= true,
xleftmargin=0ex,
xrightmargin=0ex,
aboveskip=1ex,
belowskip=1ex,
frame=trbl, %trbl
numbers=none,
float=htpb,
}

\lstnewenvironment{python}[1][]{
\lstset{style=pythonstyle, frame=trbl, belowcaptionskip=\belowtitleskip}
}{}

\newcommand{\inpyth}{\lstinline[style=inlinestyle]} %[]%

\newcounter{bla}

\journal{Computer Physics Communications}

\begin{document}

\begin{frontmatter}



\title{Oasis: a high-level/high-performance open source Navier-Stokes solver}


\author[a,b]{Mikael Mortensen\corref{author}}
\address[a]{University of Oslo, Moltke Moes vei 35, 0851 Oslo, Norway}
\address[b]{Center for Biomedical Computing at Simula Research Laboratory, P.O.Box 134, N-1325 Lysaker, Norway}
\author[b,c]{Kristian Valen-Sendstad}

\address[c]{University of Toronto, 5 Kings College Road, Toronto, ON, Canada}

\cortext[author] {Corresponding author.\\\textit{E-mail address:} mikaem@math.uio.no}

\begin{abstract}
\emph{Oasis} is a high-level/high-performance finite element Navier-Stokes solver written from scratch in Python using building blocks from the FEniCS project (fenicsproject.org). The solver is unstructured and targets large-scale applications in complex geometries on massively parallel clusters. \emph{Oasis} utilizes MPI and interfaces, through FEniCS, to the linear algebra backend PETSc. \emph{Oasis} advocates a high-level, programmable user interface through the creation of highly flexible Python modules for new problems. Through the high-level Python interface the user is placed in complete control of every aspect of the solver. A version of the solver, that is using piecewise linear elements for both velocity and pressure, is shown reproduce very well the classical, spectral, turbulent channel simulations of Moser, Kim and Mansour at $Re_{\tau}=180$ [Phys. Fluids, vol 11(4), p. 964].  The computational speed is strongly dominated by the iterative solvers provided by the linear algebra backend, which is arguably the best performance any similar implicit solver using PETSc may hope for. Higher order accuracy is also demonstrated and new solvers may be easily added within the same framework.

%A submitted program is expected to be of benefit to other physicists or physical chemists, or be an exemplar of good programming practice, or illustrate new or novel programming techniques which are of importance to some branch of computational physics or physical chemistry.
%
%Acceptable program descriptions can take different forms. The following Long Write-Up structure is a suggested structure but it is not obligatory. Actual structure will depend on the length of the program, the extent to which the algorithms or software have already been described in literature, and the detail provided in the user manual.
%
%Your manuscript and figure sources should be submitted through the Elsevier Editorial System (EES) by using the online submission tool at \\
%http://www.ees.elsevier.com/cpc.
%
%In addition to the manuscript you must supply: the program source code; job control scripts, where applicable; a README file giving the names and a brief description of all the files that make up the package and clear instructions on the installation and execution of the program; sample input and output data for at least one comprehensive test run; and, where appropriate, a user manual. These should be sent, via email as a compressed archive file, to the CPC Program Librarian at cpc@qub.ac.uk.

\end{abstract}


\end{frontmatter}

% \linenumbers

% Computer program descriptions should contain the following
% PROGRAM SUMMARY.

{\bf PROGRAM SUMMARY}
  %Delete as appropriate.

\begin{small}
\noindent
{\em Manuscript Title:}  Oasis: a high-level/high-performance open source Navier Stokes solver                                       \\
{\em Authors: } Mikael Mortensen, Kristian Valen-Sendstad   \\
{\em Program Title:} Oasis                    \\
{\em Journal Reference:}                                      \\
  %Leave blank, supplied by Elsevier.
{\em Catalogue identifier:}                                   \\
  %Leave blank, supplied by Elsevier.
{\em Licensing provisions:} GNU Lesser GPL version 3 or any later version  \\
  %enter "none" if CPC non-profit use license is sufficient.
{\em Programming language:}  Python/C++                      \\
{\em Computer:Any single laptop computer or cluster.}    \\
  %Computer(s) for which program has been designed.
{\em Operating system: Any(Linux, OSX, Windows)}                                       \\
  %Operating system(s) for which program has been designed.
{\em RAM:} a few Megabytes to several hundred Gigabytes.    \\
  %RAM in bytes required to execute program with typical data.
{\em Number of processors used:} 1 - 1000          \\
  %If more than one processor.
%{\em Supplementary material:}                                 \\
  % Fill in if necessary, otherwise leave out.
  % Please give some freely chosen keywords that we can use in a
  % cumulative keyword index.
{\em Classification:} 12                              \\
  %Classify using CPC Program Library Subject Index, see (
  % http://cpc.cs.qub.ac.uk/subjectIndex/SUBJECT_index.html)
  %e.g. 4.4 Feynman diagrams, 5 Computer Algebra.
{\em External routines/libraries:} FEniCS \\
(www.fenicsproject.org, that in turn depends on a number of external libraries like MPI, PETSc, Epetra, Boost and ParMetis)  \\
{\em Nature of problem:}\\
  %Describe the nature of the problem here.
  Incompressible, Newtonian fluid flow.
   \\
{\em Solution method:}\\
  %Describe the method solution here.
  The finite element method.
   \\
{\em Unusual features:}\\
  %Describe any unusual features of the program/problem here.
  FEniCS automatically generates and compiles low-level C++ code based on high-level Python code.
   \\
%{\em Additional comments:}\\
%  %Provide any additional comments here.
%   \\
%{\em Running time:}\\
%  %Give an indication of the typical running time here.
%   \\

\end{small}

\section{Introduction}
\label{sec:introduction}

The Navier-Stokes equations describe the flow of incompressible, Newtonian fluids. The equations are transient, nonlinear and velocity is non-trivially coupled with pressure. A lot of research has been devoted to finding efficient ways of linearizing, coupling and solving these equations. Many commercial solvers for Computational Fluid Dynamics (CFD) are available, and, due to the complexity of the high-level implementations (usually Fortran or C), users are often operating these solvers through a Graphical User Interface (GUI). To implement a generic, unstructured Navier-Stokes solver from scratch in a low-level language like C or Fortran is a considerable and time-consuming task involving tens of thousands of lines of error prone code that require much maintenance. Nowadays, as will be shown in this paper, the use of new and modern high-level software tools enables developers to cut the size of programs down to a few hundred lines and development times to hours. 

The implementation of any unstructured (Eulerian) CFD-solver requires a computational mesh. For most CFD software packages today the mesh is generated by a third-party software like, e.g., the open source projects VMTK \cite{vmtk}, Gmsh \cite{gmsh} or Cubit \cite{cubit}. To solve the governing equations on this computational mesh, the equations must be linearized and discretized such that a solution can be found for a certain (large) set of degrees of freedom. Large systems of linear equations need to be assembled and subsequently solved by appropriate direct or iterative methods. Like for mesh generation, basic linear algebra, with matrix/vector storage and operations, is nowadays most commonly outsourced to third-party software packages like PETSc \cite{petsc-web-page} and Trilinos \cite{trilinos} (see, e.g., \cite{openfvm, fluidity, oofem}). 
%Furthermore, both PETSc and Trilinos provide high-level Python interfaces (petsc4py and pytrilinos) to their low-level linear algebra solvers. This means that highly efficient C/C++ routines for computationally demanding linear algebra can be called from a scripting language without the need of firing up a compiler and at practically no additional cost.  

With both mesh generation and linear algebra outsourced, the main job of CFD solvers boils down to linearization, discretization and assembly of the linear system of equations. This is by no means a trivial task as it requires, e.g., maps from computational cells to global degrees of freedom and connectivity of cells, facets and vertices. For parallel performance it is also necessary to distribute the mesh between processors and set up for inter-communication between compute nodes. Fortunately, much of the Message Passing Interface (MPI) is already handled by the providers of basic linear algebra. When it comes down to the actual discretization, the most common approaches are probably the finite volume method, which is very popular for fluid flow, finite differences or the finite element method.

%The finite volume method, which is very popular for fluid flows, is a low-order method that uses a weak form based on integrating the governing equations over cells (or control volumes), making extensive use of the divergence theorem to ensure local (and global) conservation of transported properties like mass. The finite element method is another method that is also based on a weak variational form of the governing equations, but here the solution may be picked from a rich family of basis functions, possibly of higher order. 

FEniCS \cite{fenics} is a generic open source software framework that aims at automating the discretization of differential equations through the finite element method. FEniCS takes full advantage of specialized, reliable and robust third-party providers of computational software and interfaces to both PETSc and Trilinos for linear algebra and several third-party mesh generators. FEniCS utilizes the Unified Form Language (UFL, \cite{alnes14}) and the FEniCS Form Compiler (FFC, \cite{Kirby:2006}) to automatically generate low-level C++ code that efficiently evaluates any equation formulated as a finite element variational form. The FEniCS user has to provide the high-level variational form that is to be solved, but does not need to actually perform any coding on the level of the computational cell, or element. A choice is made of finite element basis functions, and code is then generated for the form accordingly. There is a large library of possible finite elements to choose from and they may be combined both implicitly in a coupled manner or explicitly in a segregated manner - all at the same level of complexity to the user. The user never has to see the generated low-level code, but, this being an open source project, the code is wide open for inspection and even manual fine-tuning and optimization is possible. 

%In this paper we will describe a Navier-Stokes solver written from scratch in Python, using building blocks from FEniCS. In its simplest form a complete Navier-Stokes solver for unstructured three-dimensional\footnote{There is practically no difference between the code for 2D or 3D problems; only the mesh is different.} meshes can be written using less than 50 lines of Python-code. As will be shown, though, a more efficient implementation requires a bit more effort.

In this paper we will describe the Navier-Stokes solver \emph{Oasis}, that is written from scratch in Python, using building blocks from FEniCS and the PETSc backend. Our goal with this paper is to describe a code that is (i) short and easily understood, (ii) easily configured and (iii) as fast and accurate as state-of-the-art Navier-Stokes solvers developed entirely in low-level languages.

We assume that the reader has some basic knowledge of how to write simple solvers for partial differential equations using the FEniCS framework. Otherwise, reference is given to the online FEniCS tutorial \cite{fenicstutorial}.

\section{Fractional step algorithm}
\label{sec:frac_step}
In \emph{Oasis} we are solving the incompressible Navier-Stokes equations, optionally complemented with any number of passive or reactive scalars. The governing equations are thus
\begin{align}
  \frac{\partial \bm{u}}{\partial t} + (\bm{u} \cdot \nabla) \bm{u} &= \nu \nabla^2 \bm{u} - \nabla p + \bm{f}, \label{eq:NS} \\
  \nabla \cdot \bm{u} &= 0, \label{eq:div} \\
  \frac{\partial c_{\alpha}}{\partial t} + \bm{u} \cdot \nabla c_{\alpha} &= D_{\alpha} \nabla^2 c_{\alpha} + f_{\alpha}, \label{eq:scalar} 
\end{align}
where $\bm{u}(\bm{x}, t)$ is the velocity vector, $\nu$ the kinematic viscosity, $p(\bm{x}, t)$ the fluid pressure, $c_{\alpha}(\bm{x}, t)$ is the concentration of species $\alpha$ and $D_{\alpha}$ its diffusivity. Any volumetric forces (like buoyancy) are denoted by $\bm{f}(\bm{x}, t)$ and chemical reaction rates (or other scalar sources) by $f_{\alpha}(\bm{c})$, where $\bm{c}(\bm{x}, t)$ is the vector of all species concentrations. The constant fluid density is incorporated into the pressure. Note that through the volumetric forces there is a possible feedback to the Navier-Stokes equations from the species, and, as such, a Boussinesq formulation for natural convection (see, e.g., \cite{christon02}) is possible within the current framework.

We will now outline a generic fractional step method, where the velocity and pressure are solved for in a segregated manner. Since it is important for the efficiency of the constructed solver, the velocity vector $\bm{u}$ will be split up into its individual components $u_k$.\footnote{FEniCS can alternatively solve vector equations where all components are coupled.} Time is split up into uniform intervals\footnote{It is trivial to use nonuniform intervals, but uniform is used here for convenience.} using a constant time step $\triangle t = t^n-t^{n-1}$, where superscript $n$ is an integer and $t^n \in \mathcal{R}^+$. The governing equations are discretized in both space and time. Discretization in space is performed using finite elements, whereas discretization in time is performed with finite differences.  Following Simo and Armero \cite{simo94} the generic fractional step algorithm can be written as
\begin{align}
\frac{u_k^I - {u}_k^{n-1}}{\triangle \text{t}} + B_k^{n-1/2} &= \nu \nabla^2 \tilde{{u}}_k - \nabla_k p^{*} + {f}_k^{n-1/2} \notag \\
&\text{for} \,\, k=1, \ldots, d,
  \label{eq:NStentative} \\
 \nabla^2 \varphi &= - \frac{1}{\triangle \text{t}} \nabla \cdot \bm{u}^I, \label{eq:pressure} \\
\frac{{u}_k^n - {u}_k^{I}}{\triangle \text{t}} =  - \nabla_k \varphi \quad &\text{for} \,\, k=1, \ldots, d \label{eq:correction}, \\
\frac{c_{\alpha}^n - c_{\alpha}^{n-1}}{\triangle \text{t}} + B_{\alpha}^{n-1/2} &= D_{\alpha} \nabla^2 \tilde{c}_{\alpha} + f_{\alpha}^{n-1/2}, \label{eq:scalar_disc}
\end{align}
where $u_k^n$ is component $k$ of the velocity vector at time $t^n$,  $d$ is the dimension of the problem, $ \varphi = p^{n-1/2}-p^{*}$ is a pressure correction and $p^{*}$ is a tentative pressure. We are solving for the velocity and pressure on the next time step, i.e., $u_k^n$ for $k=1,\ldots, d$ and $p^{n-1/2}$. However, the tentative velocity equation (\ref{eq:NStentative}) is solved with the tentative velocity component $u_k^I$ as unknown. To avoid strict time step restrictions, the viscous term is discretized using a semi-implicit Crank-Nicolson interpolated velocity component $\tilde{u}_k = 0.5\,({u}_k^I+{u}_k^{n-1})$. The nonlinear convection term is denoted by $B_k^{n-1/2}$, indicating that it should be evaluated at the midpoint between time steps $n$ and $n-1$. Two different discretizations of convection are currently used by \emph{Oasis}
\begin{align}
  B_{k}^{n-1/2} &= \frac{3}{2} \bm{u}^{n-1} \cdot \nabla u_k^{n-1} - \frac{1}{2} \bm{u}^{n-2} \cdot \nabla u_k^{n-2}, \label{eq:convection_ABE}\\
  B_{k}^{n-1/2} &= \overline{\bm{u}} \cdot \nabla \tilde{u}_k, \label{eq:convection_ABCN}
\end{align}
where the first is a fully explicit Adams-Bashforth discretization and the second is implicit, with an Adams-Bashforth projected convecting velocity vector $\overline{\bm{u}} = 1.5\, \bm{u}^{n-1} - 0.5\, \bm{u}^{n-2}$ and Crank-Nicolson for the convected velocity. Both discretizations are second order accurate in time, and, since the convecting velocity is known,  there is no implicit coupling between the (possibly) three velocity components solved for. 

Convection of the scalar is denoted by $B_{\alpha}^{n-1/2}$. The term must be at most linear in $c_{\alpha}^n$ and otherwise any known velocity and scalar may be used in the discretization. Note that solving for $c_{\alpha}^n$ the velocity $\bm{u}^n$ will be known and may be used to discretize $B_{\alpha}^{n-1/2}$. The discretization used in \emph{Oasis} is
\begin{align*}
  B_{\alpha}^{n-1/2} = \overline{\bm{u}} \cdot \nabla \tilde{c}_{\alpha}
\end{align*}
where $\tilde{c}_{\alpha} = 0.5\,(c_{\alpha}^n+c_{\alpha}^{n-1})$.

An iterative fractional step method involves solving Eq.~(\ref{eq:NStentative}) for all tentative velocity components and (\ref{eq:pressure}) for a pressure correction. The procedure is repeated a desired number of times before finally a velocity correction (\ref{eq:correction}) is solved to ensure conservation of mass before moving on to the next time step. The fractional step method can thus be outlined as shown in Algorithm \ref{fig:fractionalstep}. Note that if the momentum equation depends on the scalar (e.g., when using a Boussinesq model), then there may also be a second iterative loop over Navier-Stokes and temperature. The iterative scheme shown in Algorithm~\ref{fig:fractionalstep} is based on the observation that the tentative velocity computed in Eq.~(\ref{eq:NStentative}) only depends on previous known solutions $\bm{u}^{n-1}, \bm{u}^{n-2} $ and not $\bm{u}^n$. As such, the velocity update can be placed outside the inner iteration. In case of an iterative scheme where the convection depends on $\bm{u}^n$ (e.g., $\bm{u}^n\cdot \nabla \tilde{u}_k$) the update would have to be moved inside the inner loop.

\begin{algorithm}
Set time and initial conditions\\
t = 0\\
\For {\emph{time steps n = 0, 1, 2, ...}} {
  t = t + dt \\
  \For {\emph{inner iterations i = 0, 1, ...}}{
    $\varphi = p^* = p^{n-1/2}$ \\ 
    solve (\ref{eq:NStentative}) for ${u}_k^I,\, k=1,\ldots, d $\\
    solve (\ref{eq:pressure}) for $p^{n-1/2}$  \\
    $\varphi = p^{n-1/2} - \varphi$ \\
  }  
  solve (\ref{eq:correction}) for ${u}_k^n, \,k=1,\ldots, d $\\
  solve (\ref{eq:scalar_disc}) for $c_{\alpha}^{n}$\\
  update to next timestep
}
\caption{Generic fractional step algorithm for the Navier-Stokes equations. }
\label{fig:fractionalstep}
\end{algorithm}

We now have an algorithm that can be used to integrate the solution forward in time, and it is clear that the fractional step algorithm allows us to solve for the coupled velocity and pressure fields in a computationally efficient segregated manner. The efficiency and long term stability (see \cite{simo94}) are the main motivations for our choice of algorithm. However,  we should mention here that there are plenty of similar, alternative algorithms for time stepping of segregated solvers. The most common algorithm is perhaps Pressure Implicit with Splitting of Operators (PISO) \cite{piso}, which is used by both Ansys-Fluent \cite{fluent}, Star-CD \cite{starcd} and OpenFOAM \cite{openfoam}. A completely different strategy would be to solve for velocity and pressure simultaneously (coupled solvers). Using FEniCS such a coupled approach is straightforward to implement, and, in fact, it requires less coding than the segregated one. However, since the coupled approach requires more memory than a segregated, and since there are more issues with the efficiency of linear algebra solvers, the segregated approach is favoured here. 

We are still left with the spatial discretization and the actual implementation. To this end we will first show how the implementation can be performed naively, using very few lines of Python code. We will then, finally, describe the implementation of the high-performance solver.

\section{Variational formulations for the fractional step solver}
\label{sec:variational}

The governing PDEs (\ref{eq:NStentative}), (\ref{eq:pressure}), (\ref{eq:correction}) and (\ref{eq:scalar_disc}) are discretized with the finite element method in space on a bounded domain $\Omega \subset R^d$, with $2 \leq d \leq 3$, and the boundary $\partial \Omega$. Trial and test spaces for the velocity components are defined as
\begin{align}
 V &= \{ v\in H^1(\Omega): v=u_0 \,\mathrm{ on }\, \partial \Omega \}, \notag \\
 \hat{V} &= \{ v\in H^1(\Omega): v=0\, \mathrm{ on }\, \partial \Omega \},
 \label{eq:H1-spaces}
\end{align}
where $u_0$ is a prescribed velocity component on part $\partial \Omega$ of the boundary and $H^1(\Omega)$ is the Sobolev space containing functions $v$ such that $v^2$ and $|\nabla v|^2$ have finite integrals over $\Omega$.  
Both the scalars and pressure use the same $H^1(\Omega)$ space without the restricted boundary part. The  test functions for velocity component and pressure are denoted as $v$ and $q$, respectively, whereas the scalar simply uses the same test function as the velocity component.

To obtain a variational form for component $k$ of the tentative velocity vector, we multiply equation (\ref{eq:NStentative}) by $v$ and then integrate over the entire domain using integration by parts on the Laplacian
\begin{multline}
\int_{\Omega} \Big ( \frac{u_k^{I} - u_k^{n-1}}{\triangle \text{t}} + B_k^{n-1/2} \Big) v + \nu \nabla \tilde{u}_k \cdot \nabla v \, \mathrm{d}x = \\
\int_{\Omega} \Big (- \nabla_k p^{*} + f_k^{n-1/2} \Big) \, v \, \mathrm{d}x + \int_{\partial \Omega} \nu \nabla_n \tilde{u}_k \, v\, \mathrm{d}s .  \label{eq:NStentativeFEM} 
\end{multline}
Here $\nabla_n$ represents the gradient in the direction of the outward normal on the boundary. Note that the trial function $u_k^{I}$ enters also through the Crank-Nicolson velocity component $\tilde{u}_k = 0.5(u_k^{I}+u_k^{n-1})$. The boundary term is only important for some boundaries and is neglected for the rest of this paper.

The variational form for the pressure correction is obtained by multiplying Eq.~(\ref{eq:pressure})  by $q$ and then integrating over the domain, using again integration by parts
\begin{equation}
\int_{\Omega} \nabla \varphi \cdot \nabla q \, \mathrm{d}x - \int_{\partial \Omega} \nabla_n \varphi \, q \mathrm{d}s = \int_{\Omega} \frac{\nabla \cdot \bm{u}^I}{\triangle \text{t}} q \, \mathrm{d}x.  \label{eq:pressureVF} 
\end{equation}
The boundary integral can be neglected for all parts of the domain where the velocity is prescribed.

A variational form for the velocity update of component $k$ is obtained by multiplying (\ref{eq:correction}) by $v$ and integrating over the domain
\begin{equation}
 \int_{\Omega} \frac{u_k^n - {u}_k^{I}}{\triangle \text{t}} v \, \mathrm{d}x =  - \int_{\Omega} \nabla_k \varphi \, v \,\mathrm{d}x. \label{eq:correction_i_VF}
\end{equation}

Finally, a variational form for the scalar component $\alpha$ is obtained by multiplying Eq.~(\ref{eq:scalar_disc}) by $v$, and then integrating over the domain using integration by parts on the diffusion term
\begin{multline}
\int_{\Omega} \Big ( \frac{c_{\alpha}^n - c_{\alpha}^{n-1}}{\triangle \text{t}} + B_{\alpha}^{n-1/2} \Big) \, v \, + D_{\alpha} \nabla \tilde{c}_{\alpha} \cdot \nabla v\, \mathrm{d}x = \\
\int_{\Omega} f_{\alpha}^{n-1/2} \, v \, \mathrm{d}x + \int_{\partial \Omega} D_{\alpha} \nabla_n \tilde{c}_{\alpha} \, v\, \mathrm{d}s .  \label{eq:scalar_var} 
\end{multline}

\section{Oasis}
\label{sec:oasis}
We now have all the variational forms that together constitute a fractional step solver for the Navier-Stokes equations, complemented with any number of scalar fields. We will now describe how the fractional step algorithm has been implemented in \emph{Oasis} and discuss the design of the solver package. For installation of the software, see the user manual \cite{oasismanual}. Note that this paper refers to version 1.3 of the \emph{Oasis} solver, which in turn is consistent with version 1.3 of FEniCS.

\subsection{Python package}
\label{sec:pythonpackage}
\begin{figure}
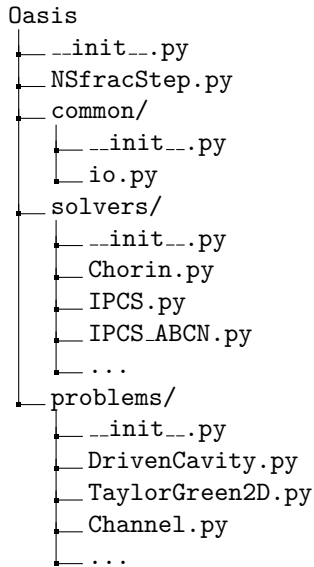

\dirtree{%
 .1 Oasis. 
 .2 {\_\_init\_\_.py}.
 .2 {NSfracStep.py}. 
 .2 common/.
 .3 {\_\_init\_\_.py}.
 .3 {io.py}.
 .2 solvers/.
 .3 {\_\_init\_\_.py}.
 .3 {Chorin.py}.
 .3 {IPCS.py}.
 .3 {IPCS\_ABCN.py}.
 .3 {...}.
 .2 problems/.
 .3 {\_\_init\_\_.py}. 
 .3 {DrivenCavity.py}. 
 .3 {TaylorGreen2D.py}.
 .3 {Channel.py}. 
 .3 {...}.
}
\caption{Directory tree structure of Python package \emph{Oasis}. }
\label{fig:directorytree}
\end{figure}

The \emph{Oasis} solver is designed as a Python package with tree structure shown in Fig.~\ref{fig:directorytree}. 
The generic fractional step algorithm is implemented in the top level Python module \inpyth{NSfracStep.py}{} and the solver is run by executing this module within a Python shell using appropriate keyword arguments, e.g.,
\vskip 1ex
\begin{python}
>>> python NSfracStep.py problem=Channel solver=IPCS
\end{python}
\vskip 1ex

\noindent The fractional step solver pulls in a required mesh, parameters and functions from two submodules located in folders \inpyth{solvers}{} and \inpyth{problems}{}. The user communicates with the solver through the implementation of new problem modules in the \inpyth{problems}{} folder. With the design choice of placing the solver at the root level of a Python module, there is a conscious decision of avoiding object oriented classes. However, remembering that everything in Python is an object, we still, as will be shown, make heavy use of overloading Python objects (functions, variables). 

The fractional step module \inpyth{NSfracStep.py}{} is merely one hundred lines of code (excluding comments and spaces) dedicated to allocation of necessary storage and variables, plus the implementation of the generic fractional step Algorithm \ref{fig:fractionalstep}. The first half of \inpyth{NSfracStep.py}{} is shown in Fig.~\ref{lst:allocation}. Except from the fact that most details are kept in submodules, the design is very similar to most FEniCS Python demos, and, as such, \emph{Oasis} should feel familiar and be quite easily accessible to new users with some FEniCS experience.

\begin{figure*}
\begin{python}
from common import *

commandline_kwargs = parse_command_line()

# Get the problem from commandline
problem = commandline_kwargs.get("problem", "DrivenCavity")

# import mesh, NS_parameters, body_force, create_bcs, velocity_degree, etc...
exec("from problems.{} import *".format(problem))

# Update NS_parameters with parameters modified through the command line 
NS_parameters.update(commandline_kwargs)
vars().update(NS_parameters)  

# Import functionality from chosen solver
exec("from solvers.{} import *".format(solver))

# Declare function spaces and trial and test functions
V = FunctionSpace(mesh, "Lagrange", velocity_degree)
Q = FunctionSpace(mesh, "Lagrange", pressure_degree)
u, v = TrialFunction(V), TestFunction(V)
p, q = TrialFunction(Q), TestFunction(Q)

# Get dimension of problem
dim = mesh.geometry().dim()

# Create list of components we are solving for
u_components = map(lambda x: "u"+str(x), range(dim))  # velocity components
uc_comp  =  u_components + scalar_components          # velocity + scalars
sys_comp = u_components + ["p"] + scalar_components   # velocity + pressure + scalars

# Create dictionaries for the solutions at three timesteps
q_  = {ui: Function(V) for ui in uc_comp}
q_1 = {ui: Function(V) for ui in uc_comp}
q_2 = {ui: Function(V) for ui in u_components} # Note only velocity

# Allocate solution for pressure field and correction
p_ = q_["p"] = Function(Q)
phi_ = Function(Q)

# Create vector views of the segregated velocity components    
u_  = as_vector([q_ [ui] for ui in u_components]) # Velocity vector at t
u_1 = as_vector([q_1[ui] for ui in u_components]) # Velocity vector at t - dt
u_2 = as_vector([q_2[ui] for ui in u_components]) # Velocity vector at t - 2*dt

# Set kinematic viscosity constant
nu = Constant(NS_parameters["nu"])

# Set body force
f = body_force(**vars())

# Initialize solution
initialize(**vars())

# Get boundary conditions
bcs = create_bcs(**vars())

\end{python}
\caption{The opening section of \inpyth{NSfracStep.py}. Allocation of necessary storage and parameters for solving the momentum equation through its segregated components. Note that a mesh, some parameters (for e.g., viscosity, time step, end time etc), and some functions (for e.g., body force, boundary conditions or initializing the solution)  must be imported from the problem module. The UFL function \inpyth{as_vector} creates vectors (\inpyth{u_, u_1, u_2}) from the segregated velocity components. The built-in function \inpyth{vars()} returns the current moduoles namespace. Neglecting scalar components the list \inpyth{sys_comp = ["u0", "u1", "p"]} for 2D and \inpyth{["u0", "u1", "u2", "p"]} for 3D problems. The list is used as keys for the dictionary \inpyth{bcs}. }
\label{lst:allocation}
\end{figure*}

Consider the three functions towards the end of Fig.~\ref{lst:allocation} that take \inpyth{**vars()} as argument. The \inpyth{body_force} function returns $\bm{f}$ in (\ref{eq:NS}) and should thus by default return a \inpyth{Constant} vector of zero values (length 2 or 3 depending on whether the problem is 2D or 3D). The \inpyth{initialize} function initializes the solution in \inpyth{q_, q_1, q_2} and \inpyth{create_bcs} must return a dictionary of boundary conditions. These functions are clearly problem specific and thus default implementations are found in the \inpyth{problems/__init__.py} module that all new problems are required to import from. The default functions may then be overloaded as required by the user in the new problem module (see, e.g., Fig.~\ref{fig:drivencavity}). An interesting feature is the argument \inpyth{**vars()}, which is used for all three functions. The Python built-in function \inpyth{vars()} returns a dictionary of the current module's namespace, i.e., it is here \inpyth{NSfracStep}'s namespace containing \inpyth{V, Q, u, v, } and all the other variables seen in Fig.~\ref{lst:allocation}. When \inpyth{**vars()} is used in a function's signature, any variable declared within \inpyth{NSfracStep}'s namespace may be unpacked in that function's list of arguments and accessed by reference. Figure~\ref{lst:defaulthooks} illustrates this nicely through the default implementations (found in \inpyth{problems/__init__.py}) of the three previously mentioned functions. 
%Note how any variable required by the function is simply unpacked in the list of arguments, like for instance \inpyth{mesh}{} in \inpyth{body_force}. All other variables from the \inpyth{NSfracStep}'s namespace are soaked up by the trailing \inpyth{NS_namespace} dictionary. Since the calling uses a reference to the namespace, there is very little overhead, yet unlimited flexibility, in calling functions this way.
\begin{figure}
\begin{python}
def body_force(mesh, **NS_namespace):
  """Specify body force"""
  dim = mesh.geometry().dim()
  return Constant((0,)*dim)
    
def initialize(**NS_namespace):
  """Initialize solution. """
  pass

def create_bcs(sys_comp, **NS_namespace):
  """Return dictionary of Dirichlet 
  boundary conditions."""
  return {ui: [] for ui in sys_comp}
\end{python}
\caption{Default implementations of three of the functions found in \inpyth{problems/__init__.py}{}. }
\label{lst:defaulthooks}
\end{figure}

After initialization the solution needs to be advanced in time. The entire implementation of the time integration performed in \inpyth{NSfracStep.py} is shown in Fig.~\ref{fig:timeloop}, that closely resembles Algorithm~\ref{fig:fractionalstep}. In Fig.~\ref{fig:timeloop} the functions ending in \inpyth{hook}{} are imported through the \inpyth{problems} submodule, \inpyth{save_solution} from \inpyth{common} and the rest of the functions are imported from the \inpyth{solvers} submodule. 
\begin{figure}[ht!]
\begin{python}
# Preassemble and prepare solver
vars().update(setup(**vars()))

# Enter loop for time advancement
while t < T and not stop:
  t += dt
  inner_iter = 0
  # Do something at start of timestep    
  start_timestep_hook(**vars())
    
  # Enter velocity/pressure inner loop
  for inner_iter < max_iters:
    inner_iter += 1        
    if inner_iter == 1:
      assemble_first_inner_iter(**vars())
         
    # Solve Eq. (17)
    for i, ui in enumerate(u_components):
      velocity_tentative_assemble(**vars())
      velocity_tentative_hook    (**vars())
      velocity_tentative_solve   (**vars())
            
    # Solve Eq. (18)
    pressure_assemble(**vars())
    pressure_hook    (**vars())
    pressure_solve   (**vars())
         
  # Solve Eq. (19)
  velocity_update(**vars())
        
  # Solve for all scalar components (20)
  if len(scalar_components) > 0:
    scalar_assemble(**vars())
    for ci in scalar_components:    
      scalar_hook (**vars())
      scalar_solve(**vars())
        
  # Do something at end of timestep
  temporal_hook(**vars())
    
  # Save and update to next timestep
  stop = save_solution(**vars())

# Finalize solver
theend_hook(**vars())

\end{python}
\caption{Time loop in \inpyth{NSfracStep.py}{}}
\label{fig:timeloop}
\end{figure}

The \inpyth{common} submodule basically contains routines for parsing the command line and for storing and retrieving the solution (\inpyth{common/io.py}{}). There is, for example, a routine here that can be used if the solver needs to be restarted from a previous simulation. The \inpyth{problems} and \inpyth{solvers} submodules are more elaborate and will be described next.

\subsubsection*{The problems submodule}

\emph{Oasis} is a programmable solver and the user is required to implement the problem that is to be solved. The implemented problem module's namespace must include at least a computational mesh and functions for specifying boundary conditions and initialization of the solution. Other than that, the user may interact with \inpyth{NSfracStep} through certain \inpyth{hook} files strategically placed within the time advancement loop, as seen in Fig.~\ref{fig:timeloop}, and as such there is no need to modify \inpyth{NSfracStep} itself. 

Consider a lid driven cavity with $\Omega = [0, 1]\times[0, 1]$. The velocity boundary conditions are $\bm{u} = (1, 0)$ for the top lid ($y=1$) and zero for the remaining walls. We start the simulations from a fluid at rest and advance the solution in time steps of $\triangle t=0.001$ from $t=0$ to $t=1$. The viscosity is set to $\nu=0.001$. This problem can be implemented as shown in Fig.~\ref{fig:drivencavity}. Here we have made use of the standard python package \inpyth{numpy} and two \inpyth{dolfin} classes \inpyth{UnitSquareMesh} and \inpyth{DirichletBC}. \inpyth{UnitSquareMesh} creates a computational mesh on the unit square, whereas   \inpyth{DirichletBC} creates Dirichlet boundary conditions for certain segments of the boundary identified through two strings \inpyth{noslip} and \inpyth{top} (\inpyth{x[0]} and \inpyth{x[1]} represent coordinates $x$ and $y$ respectively). A default set of problem parameters can be found in the dictionary \inpyth{NS_parameters} declared in \inpyth{problems/__init__.py}, and all these parameters may be overloaded, either as shown in Fig.\ref{fig:drivencavity}, or through the command line. 

A comprehensive list of parameters and their use is given in the user manual. We use preconditioned iterative Krylov solvers (\inpyth{NS_parameters["use_krylov_solvers"]=True}), and not the default direct solvers based on LU decomposition, since the former here are faster and require less memory (the exact choice of iterative solvers is discussed further in Sec\ref{sec:conclusions}). Note that FEniCS interfaces to a wide range of different linear algebra solvers and preconditioners. The iterative solvers used by \emph{Oasis} are defined in function \inpyth{get_solvers} imported from the \inpyth{solvers} submodule.
\begin{figure}[ht!]
\begin{python}
from problems import *
from numpy import cos, pi

# Create a mesh skewed towards walls
def mesh(Nx, Ny, **params):
  m = UnitSquareMesh(Nx, Ny)
  x = m.coordinates()
  x[:] = (x-0.5)*2.
  x[:] = 0.5*(cos(pi*(x-1.)/2.)+1.)
  return m

# Override some problem specific parameters
NS_parameters.update(
  nu = 0.001,
  T  = 1.0,
  dt = 0.001,
  Nx = 50,
  Ny = 50,
  use_krylov_solvers = True)

# Specify boundary conditions
noslip="std::abs(x[0]*x[1]*(1-x[0]))<1e-8"
top   ="std::abs(x[1]-1) < 1e-8"
def create_bcs(V, **NS_namespace):
  bc0  = DirichletBC(V, 0, noslip)
  bc00 = DirichletBC(V, 1, top)
  bc01 = DirichletBC(V, 0, top)
  return dict(u0 = [bc00, bc0],
              u1 = [bc01, bc0],
              p  = [])
                
# Initialize by enforcing boundary cond.
def initialize(q_1, q_2, bcs, **NS_namesp):
  for ui in q_2:
    for bc in bcs[ui]:
      bc.apply(q_1[ui].vector())
      bc.apply(q_2[ui].vector())

\end{python}
\caption{Drivencavity.py - Implementation of the driven cavity problem.}
\label{fig:drivencavity}
\end{figure}
  
To run the solver for the driven cavity problem we need to specify this through the command line - along with any other parameter we wish to modify at runtime. For example, the default size of the computational mesh has been implemented in Fig.~\ref{fig:drivencavity} as \inpyth{Nx=Ny=50}. This may be overloaded through the command line while running the solver, like
\vskip 1ex
\noindent
\begin{minipage}{\columnwidth}
\begin{python}
>>> python NSfracStep.py problem=DrivenCavity Nx=20 Ny=20
\end{python}
\end{minipage}
\vskip 1ex
\noindent The ability to overload parameters through the command line is useful for, e.g., fast convergence testing. 

The computational mesh has to be part of the problem module's namespace. However, it does not need to be defined as a callable function, like that used in Fig.~\ref{fig:drivencavity}. Three equally valid examples are

\vskip 1ex
\noindent
\begin{minipage}{\columnwidth}
\begin{python}
mesh = UnitSquareMesh(10, 10)
mesh = Mesh("SomeMesh.xml.gz")
def mesh(N, **params):
    return UnitSquareMesh(N, N)
\end{python}
\end{minipage}
\vskip 1ex
\noindent The first mesh is hardcoded in the module and cannot be modified through the commandline. The second approach, \inpyth{mesh = Mesh("some_mesh.xml.gz")}{}, is usually used whenever the mesh has been created by an external software. The third option uses a callable function, making it possible to modify the mesh size through the command line.

A complete list of all default functions and parameters that may be overloaded by the user in their implemented problem module is found in \inpyth{problems/__init__.py}. 

%A flow chart illustrating the structure of the solver is shown in Fig.~(\ref{fig:flowchart}). Here each arrow indicates a \inpyth{from module import *}{} command. That is, all dolfin's functionality is imported by the \inpyth{problems/__init__.py}{} module, where a number of other default functions are defined as well. These are then possibly overloaded by the user. The entire purpose of the hierarchical structure of modules is for the solver to be able to pull in functions as requested by the user - and to allow the user to have the last saying, by enabling the user to overload any default behaviour in the problem module.
%
%\begin{figure}
%  \includegraphics[scale=0.6]{Drawing11.png}
%  \caption{Flow chart for solver. Each arrow indicates a \inpyth{from module import *}{} command. That is, all dolfin's functionality is imported by the default\_hooks module, where a number of other defalt functions are defined as well. These are then possibly overloaded by the user in the problem module. }
%  \label{fig:flowchart}
%\end{figure}

\subsubsection*{The solvers submodule}

The finer details of the fractional step solver are implemented in the \inpyth{solvers} submodule. A list of all functions that are imported by \inpyth{NSfracStep} is found in the \inpyth{solvers/__init__.py} module. The most important can be seen in Fig.~\ref{fig:timeloop}. Note the special calling routine for the function \inpyth{setup}
\begin{python}
vars().update(setup(**vars()))
\end{python}
The purpose of this \inpyth{setup} function is to prepare the solver for time advancement. This could mean either defining UFL forms of the variational problems (see Fig.~\ref{lst:momentumNaive}) or to preassemble matrices that do not change in time, e.g., diffusion and mass matrices (see Sec.~\ref{sec:hpc}). The \inpyth{setup} function returns a dictionary and this dictionary is updated and made part of the \inpyth{NSfracStep} namespace through the use of \inpyth{vars().update}.

We may now take the naive approach and implement all variational forms exactly as described in Sec.~\ref{sec:variational}. A smart approach, on the other hand, will take advantage of certain special features of the Navier-Stokes equations. The starting point for implementing a new solver, though, will usually be the naive approach. A naive implementation requires very few lines of code, it is easy to debug and as such it can be very useful for verification of the slightly more complex and optimized solvers to be discussed in the next section.
\begin{figure}[ht!]
\begin{python}
def setup(u_, u_1, q_, q_1, u, v, p, q, 
          nu, dt, p_, f, u_components, 
          phi_, **NS_namespace):
  F, Fu = {}, {}
  U_AB = 1.5*u_1 - 0.5*u_2
  for i, ui in enumerate(u_components):
    # Crank-Nicolson velocity  
    U_CN = 0.5*(u+q_1[ui])
    
    # Tentative velocity variational form
    F[ui] = (1./dt*inner(u-q_1[ui], v)*dx
      + inner(dot(U_AB, grad(U_CN)), v)*dx
      + nu*inner(grad(U_CN), grad(v))*dx
      + inner(p_.dx(i), v)*dx 
      - inner(f[i], v)*dx)
            
    # Velocity update variational form
    Fu[ui]= (inner(u, v)*dx 
            - inner(q_[ui], v)*dx 
            + dt*inner(phi_.dx(i), v)*dx)

  # Variational form for pressure
  phi = p - p_
  Fp = (inner(grad(q), grad(phi))*dx 
     - (1./dt)*div(u_)*q*dx) 
  
  return dict(F=F, Fu=Fu, Fp=Fp)
\end{python}
\caption{Naive implementation in \inpyth{solvers/IPCS.py} of variational forms used for solving the momentum equation (\ref{eq:NStentativeFEM}), pressure correction (\ref{eq:pressureVF}) and momentum update (\ref{eq:correction_i_VF}).}
\label{lst:momentumNaive}
\end{figure}

The \inpyth{solvers/IPCS.py} module contains a naive implementation of the variational forms (\ref{eq:NStentativeFEM}), (\ref{eq:pressureVF}) and (\ref{eq:correction_i_VF}). The forms are implemented using the \inpyth{setup} function shown in Fig.~\ref{lst:momentumNaive}. Dictionaries are used to hold the forms for the velocity components, whereas there is only one form required for the pressure. Note the very close correspondence between the high-level Python code and the mathematical description of the variational forms. The variational forms are assembled and solved through the very compact routines \inpyth{velocity_tentative_solve, pressure_solve} and \inpyth{velocity_update} that are implemented as shown in Fig.~\ref{fig:ipcs_solve}. The remaining default functions are left to do nothing, as implemented already in \inpyth{solvers/__init__.py}, and as such these 4 functions shown in Figs.~\ref{lst:momentumNaive} and \ref{fig:ipcs_solve} are all it takes to complete the implementation of the naive incremental pressure correction solver. Note that this implementation works for any order of the velocity/pressure function spaces. There is simply no additional implementation cost for using higher order elements.
\begin{figure}[ht!]
\begin{python}
def velocity_tentative_solve(ui, F, q_,bcs, 
                           **NS_namespace):
  A, L = system(F[ui])
  solve(A == L, q_[ui], bcs[ui])

def pressure_solve(Fp, p_, bcs, phi_, 
                   **NS_namespace):
  # Compute pressure 
  phi_.vector()[:] = p_.vector()
  A, L = system(F[ui])
  solve(A == L, p_, bcs['p'])   

  # Normalize pressure if no bcs['p']  
  if bcs['p'] == []:
    normalize(p_.vector())
    
  # Compute correction
  phi_.vector()[:] = p_.vector() - phi_.vector()

def velocity_update(u_components, q_, bcs, 
                    Fu, **NS_namespace):
  for ui in u_components:
    A, L = system(F[ui])
    solve(A == L, q_[ui], bcs[ui])

\end{python}
\caption{Implementation in \inpyth{solvers/IPCS.py} of routines called in Fig.~\ref{fig:timeloop}.}
\label{fig:ipcs_solve}
\end{figure}

\section{High-performance implementation}
\label{sec:hpc}
The naive solver described in the previous section is very easy to implement and understand, but for obvious reasons it is not very fast. For example, the entire coefficient matrix is reassembled each timestep (see Fig.~\ref{fig:timeloop}), even though it is only the convection term that changes in time. We will now explain how the same incremental pressure correction solver can be implemented efficiently, at the cost of loosing some intuitiveness. The implementation of the high-performance solver described in this section can be found in \inpyth{solvers/IPCS_ABCN.py}. 

The most significant steps in the optimization can roughly be split into four contributions: (i) preassembling of constant matrices making up the variational forms, (ii) efficient assembly of the entire coefficient matrix, where in an intermediate form it is used also to compute large parts of the linear right hand side, (iii) use of constructed (constant) matrices for assembling terms on right hand side through fast matrix vector products and (iv) efficient use and re-use of iterative solvers with preconditioners.

To implement an efficient solver we need to split up the variational forms (\ref{eq:NStentativeFEM}), (\ref{eq:pressureVF}) and (\ref{eq:correction_i_VF}) term by term and view the equations on an algebraic level. The finite element solution, which is the product of the solver, is then written as
\begin{equation}
 u_k^I=\sum_{j=1}^{N_u} \mathcal{U}_j^{k,I}\, \phi_j,
 \label{eq:u_function}
\end{equation}
where $\phi_j$ are the basis functions and $\{\mathcal{U}_j^{k,I}\}_{j=1}^{N_u}$ are the $N_u$ degrees of freedom.

We start by inserting for the tentative velocity $u_k^I$ and $v=\phi_i$ in the bilinear terms of the variational form (\ref{eq:NStentativeFEM})
\begin{align}
 \int_{\Omega} u_k^I v \,dx &= \sum_{j=1}^{N_u}\left(\int_{\Omega} \phi_j\, \phi_i \, dx \right) \mathcal{U}_j^{k,I}, \\
  \int_{\Omega} \nabla u_k^I \cdot \nabla v \,dx &= \sum_{j=1}^{N_u}\left(\int_{\Omega} \nabla \phi_j \cdot \nabla \phi_i \, dx \right) \mathcal{U}_j^{k,I}.
\end{align}
Each term inside the parenthesis on the right hand side represents a matrix
\begin{align}
 M_{ij} &= \int_{\Omega} \phi_j\, \phi_i \, dx, \\
 K_{ij} &= \int_{\Omega} \nabla \phi_j \cdot \nabla \phi_i \, dx. \end{align}
The two matrices are independent of time and can be preassembled once through (\inpyth{u, v} are trial and test functions respectively)

\vskip 1ex
\noindent
\begin{minipage}{\columnwidth}
\begin{python}[b!]
M = assemble(inner(u, v)*dx)
K = assemble(inner(grad(u), grad(v))*dx)
\end{python}
\end{minipage}
\vskip 1ex

\noindent Note that the solution vectors and matrices represent the major cost in terms of memory use for the solver. The matrices are sparse and allocated by the linear algebra backend, using appropriate wrappers that are hidden to the user. The allocation takes place just once, when the matrices/vectors are created.

The nonlinear convection form contains the evolving solution and requires special attention. We use the implicit convection form given in Eq.~(\ref{eq:convection_ABCN}) and write out the implicit Crank-Nicolson convected velocity for component $k$
\begin{equation}
\overline{\bm{u}} \cdot \nabla \tilde{u}_k = 
\frac{1}{2}\, \overline{\bm{u}} \cdot \nabla \left(u_k^I + u_k^{n-1} \right).
\end{equation}
Inserting for the algebraic form of the finite element trial  and test functions, the variational form for the bilinear convection term becomes
\begin{equation}
  \int_{\Omega} \overline{\bm{u}} \cdot \nabla u_k^I \, v\, dx = \sum_{j=1}^{N_u}\left( \int_{\Omega} \overline{\bm{u}} \cdot \nabla \phi_j  \, \phi_i \, dx \right) \mathcal{U}_j^{k,I},
\end{equation}
where $\overline{\bm{u}} = 1.5\, \bm{u}^{n-1} - 0.5\, \bm{u}^{n-2}$. The convection matrix can be recognized as the term inside the parenthesis
\begin{equation}
C_{ij}^{n-1/2} = \int_{\Omega} \overline{\bm{u}} \cdot \nabla \phi_j  \, \phi_i \, dx.
\end{equation}
The convecting velocity is time-dependent and interpolated at $t^{n-1/2}$. As such, the convection matrix is also evaluated at $n-1/2$ and needs to be reassembled each timestep. To simplify notations, though, we have for the rest of this paper omitted the time notation on $C_{ij}$. The assembly of the $C_{ij}$ matrix is prepared in the \inpyth{setup} function:
\vskip 1ex
\noindent
\begin{minipage}{\columnwidth}
\begin{python}
# Defined in setup
u_ab = as_vector([Function(V) for i in           
               range(len(u_components))])
aconv = inner(v, dot(grad(u), u_ab))*dx
\end{python}
\end{minipage}
\vskip 1ex
\noindent where \inpyth{u_ab} is used as a container for the convecting velocity $\overline{\bm{u}}$. Note that \inpyth{u_ab},is assembled (see Fig~\ref{fig:assemble_first_inner_iter}) before assembling the matrix $C_{ij}$, because this leads to code that is a factor 2 faster than simply using a form based on the velocity functions at the two previous levels directly (i.e., \inpyth{aconv = inner(v, dot(grad(u), 1.5*u_1 - 0.5*u_2))*dx}).

Consider now the linear terms, where the known solution function is written as $u_k^{n-1} = \sum_{j=1}^{N_u} \mathcal{U}_j^{k, n-1} \phi_j$, where $\mathcal{U}_j^{k, n-1}$ are the known coefficients of velocity component $k$ at the previous time step $t^{n-1}$. We have the following linear terms in Eq.~(\ref{eq:NStentativeFEM}) 
\begin{align}
 \int_{\Omega} u_k^{n-1}v \,dx &= M_{ij} \, \mathcal{U}_j^{k,n-1},\\
  \int_{\Omega} \nabla u_k^{n-1} \cdot \nabla v \,dx &= K_{ij}\,\mathcal{U}_j^{k, n-1}, \\
  \int_{\Omega} \overline{\bm{u}} \cdot \nabla u_k^{n-1} \, v\, dx &= C_{ij}\, \mathcal{U}_j^{k, n-1},
\end{align}
that are all very quickly computed using simple matrix vector products.

We may now reformulate our variational problem on the algebraic level using the three assembled matrices. It is required that for each test function $v=\phi_i, i=1, \ldots, N_u$, the following equations must hold 
\begin{multline}
 \frac{M_{ij}\left(\mathcal{U}_j^{k,I} - \mathcal{U}_j^{k, n-1}\right)}{\triangle t} + \frac{C_{ij}\left(\mathcal{U}_j^{k,I}+\mathcal{U}_j^{k, n-1}\right)}{2} \\ + \nu \frac{K_{ij}\left(\mathcal{U}_j^{k,I} +\mathcal{U}_j^{k, n-1}\right)}{2} = \Phi_i^{k, n-1/2},
\end{multline}
where
\begin{equation}
\Phi_i^{k, n-1/2} = \int_{\Omega} \Big (- \nabla_k p^{*} + f_k^{n-1/2} \Big) \, \phi_i \, \mathrm{d}x.
\end{equation}
If separated into bilinear and linear terms, the following system of algebraic equations is obtained
\begin{multline}
 \left(\frac{M_{ij}}{\triangle t} + \frac{C_{ij}}{2} + \nu\frac{K_{ij}}{2}\right) \mathcal{U}_j^{k,I} = \\
 \left(\frac{M_{ij}}{\triangle t} - \frac{C_{ij}}{2} - \nu\frac{K_{ij}}{2}\right) \mathcal{U}_j^{k, n-1}+\Phi_i^{k, n-1/2}. \label{eq:tent_algebraic}
\end{multline}
If now $A_{ij}= M_{ij}/\triangle t + C_{ij}/2 + \nu K_{ij}/2$ is used as the final coefficient matrix, then the equation may be written as
\begin{equation}
  A_{ij} \, \mathcal{U}_j^{k,I} = \left(\frac{2M_{ij}}{\triangle t} -A_{ij}\right)\, \mathcal{U}_j^{k,n-1} + \Phi_i^{k, n-1/2}, \label{eq:tent_intermediate1}
\end{equation}
or 
\begin{equation}
  A_{ij} \, \mathcal{U}_j^{k,I} = b_i^{k, n-1/2}, \quad \text{for}\, k= 1, \ldots, d, \label{eq:Au=b}
\end{equation}
where $b_i^{k, n-1/2}$ is the right hand side of (\ref{eq:tent_intermediate1}). Note that the same coefficient matrix is used by all velocity components, even when there are Dirichlet boundary conditions applied. 

An efficient algorithm (\ref{alg:A}) can now be designed to assemble both large parts of the right hand side and the left hand side of Eq.~(\ref{eq:Au=b}) at the same time.
\begin{algorithm}
\begin{align}
        \text{Assemble}\,\,\, A_{ij} &\longleftarrow C_{ij} \notag \\
        A_{ij} &= M_{ij}/dt - A_{ij}/2-\nu K_{ij}/2 \notag \\      
        b_i^{k, n-1/2} &= f_i^{k, n-1/2} + A_{ij}\,\mathcal{U}_j^{k, n-1}, \notag \\
        & \quad\quad\quad\quad \text{for}\quad k=1,\ldots, d \notag \\
        A_{ij} &= -A_{ij} + 2M_{ij}/dt \notag
\end{align}
\caption{Efficient algorithm for assembling the coefficient matrix $A_{ij}$, where most of the right hand side of Eq.~(\ref{eq:Au=b}) is assembled in an intermediate step. }
\label{alg:A}
\end{algorithm}

Algorithm (\ref{alg:A}) is implemented as shown in Fig.~\ref{fig:assemble_first_inner_iter}. At the end of this algorithm, most of $b^{k, n-1/2}$ (except from the pressure gradient) has been assembled and the coefficient matrix $A_{ij}$ is ready to be used in Eq.~(\ref{eq:Au=b}). The convection matrix needs to be reassembled each new time step, but only on the first inner velocity pressure iteration since $\overline{\bm{u}}$ only contains old and known velocities, not the new $u_k^{n}$. For this reason the code in Fig.~\ref{fig:assemble_first_inner_iter} is placed inside  \inpyth{assemble_first_inner_iter}, called in Fig.~\ref{fig:timeloop}. Notice that there is no separate matrix used for $2M_{ij}/\triangle t - A_{ij}$ or $C_{ij}$ and the total memory cost of the algorithm is exactly three individual sparse matrices ($A_{ij}, M_{ij}$ and $K_{ij}$). The sparsity pattern of the matrices is computed on the first assemble and the matrix axpy operations take advantage of the fact that all these matrices share the same pattern. 
\begin{figure}[t!]
\begin{python}
# assemble convecting velocity
for i, ui in enumerate(u_components):
  u_ab[i].vector().zero()
  u_ab[i].vector().axpy(1.5, x_1[ui])
  u_ab[i].vector().axpy(-0.5, x_2[ui])

# assemble convection into A
A = assemble(a_conv, tensor=A, 
             reset_sparsity=False) 
  
# Negative convection on the rhs 
A._scale(-0.5)            
  
# Add mass and diffusion matrix
A.axpy(1./dt, M, True)
A.axpy(-0.5*nu, K, True) 
  
# Compute parts of rhs vector
for ui in u_components:
  b_tmp[ui].zero()
  # Add body force stored in b0
  b_tmp[ui].axpy(1., b0[ui])
  # Add transient, convection and diffusion
  b_tmp[ui].axpy(1., A*x_1[ui])
        
# Reset matrix for lhs
A._scale(-1.)
A.axpy(2./dt, M, True)
  
# Apply boundary conditions
[bc.apply(A) for bc in bcs['u0']]
  
\end{python}
\caption{Inside \inpyth{assemble_first_inner_iter}. Fast assemble of coefficient matrix and parts of right hand side vector. A temporary rhs vector \inpyth{b_tmp} is used for each velocity component since this routine is called only on the first inner iteration. \inpyth{x_1} is the vector of degrees of freedom at $t^{n-1}$.}
\label{fig:assemble_first_inner_iter}
\end{figure}

The linear term $\Phi_i^{k, n-1/2}$ needs some further comments. Neglecting the constant forcing, $\bm{f}$, the second part of $\Phi_i^{k, n-1/2}$ is
\begin{equation}
\int_{\Omega} - \nabla_k p^{*} \, \phi_i \, \mathrm{d}x,
\end{equation}
where $p^{*}=\sum_{j=1}^{N_p} \mathcal{P}_j^{*} \hat{\phi}_j $, $\hat{\phi}_j$ is the basis function for the pressure and $\mathcal{P}_j^{*}$ are the known degrees of freedom. On algebraic form we get
\begin{align}
\int_{\Omega} \nabla_k p^{*} \, \phi_i \, \mathrm{d}x &=\sum_{j=1}^{N_p} \left(\int_{\Omega} \nabla_k \hat{\phi}_j\, \phi_i\, \mathrm{d}x \right) \mathcal{P}_j^{*}, \notag\\
&= dP_{ij}^k \,\mathcal{P}_j^{*}, \label{eq:dp_matrix}
\end{align}
where $dP_{ij}^k$ for $k=1,\ldots, d$ are $d$ matrices that are constant in time. Since the matrices can be preassembled, the computation of  $\Phi_i^{k, n-1/2}$ through a matrix vector product is very fast. Unfortunately, though, three additional matrices require storage (in 3D), which may be too expensive. In that case there is a parameter in \emph{Oasis} that can be used. Setting
\vskip 1ex
\noindent
\begin{minipage}{\columnwidth}
\begin{python}
NS_parameters["low_memory_version"] = False
\end{python}
\end{minipage}
\vskip 1ex
\noindent enables the creation of the matrices $dP_{ij}^k$. If disabled the term is computed simply through
\vskip 1ex
\noindent
\begin{minipage}{\columnwidth}
\begin{python}
assemble(inner(p_.dx(k), v)*dx)
\end{python}
\end{minipage}
\vskip 1ex
\noindent for $k=0, \ldots, d-1$. The pressure gradient is added to $b^k$ in \inpyth{velocity_tentative_assemble} and not in Fig.~\ref{fig:assemble_first_inner_iter}, since the pressure is modified on inner iterations.

The pressure correction equation can also be optimized on the algebraic level. Using trial function $p^{n-1/2}=\sum_{j=1}^{N_p} \mathcal{P}_j^{n-1/2} \hat{\phi}_j $ and test function $q=\hat{\phi}_i$ we can write (\ref{eq:pressureVF}) for each test function
%\begin{multline}
%\sum_{j=1}^{N_p} \left(\int_{\Omega} \nabla \hat{\phi}_j \nabla \hat{\phi}_i \mathrm{d}x \right) \mathcal{P}_j^{n-1/2} = \\ \sum_{j=1}^{N_p} \left(\int_{\Omega} \nabla \hat{\phi}_j \nabla \hat{\phi}_i \mathrm{d}x \right) \mathcal{P}_j^{*} \\- \int_{\Omega} \frac{\nabla \cdot \bm{u}^{I}}{\triangle \text{t}} \hat{\phi}_i\, \mathrm{d}x ,
%\end{multline}
%or
\begin{equation}
 \hat{K}_{ij}\mathcal{P}_j^{n-1/2} = \hat{K}_{ij}\mathcal{P}_j^{*}- \int_{\Omega} \frac{\nabla \cdot \bm{u}^{I}}{\triangle \text{t}} \hat{\phi}_i\, \mathrm{d}x.
\end{equation}
The Laplacian matrix $\hat{K}_{ij}$ can be preassembled. If the pressure function space is the same as the velocity function space, then $\hat{K}_{ij} = K_{ij}$ and no additional work is required. The divergence term may be computed as
\begin{align}
\int_{\Omega} \frac{\nabla \cdot \bm{u}^{I}}{\triangle \text{t}} \hat{\phi}_i\, \mathrm{d}x &= \frac{1}{\triangle \text{t}}\sum_{k=1}^{d} \left( \sum_{j=1}^{N_u} \int_{\Omega} \nabla_k \phi_j \hat{\phi}_i\, \mathrm{d}x\, \mathcal{U}_j^{k, I} \right), \notag \\
&= \frac{1}{\triangle \text{t}}\sum_{k=1}^{d}  d\mathcal{U}_{ij}^{k}\, \mathcal{U}_j^{k, I} ,
\end{align} 
where the matrices $d\mathcal{U}_{ij}^k$ for $k=1, \ldots, d$ can be preassembled. Again, the cost is three additional sparse  matrices, unless the function spaces of pressure and velocity are the same. In that case $d\mathcal{U}_{ij}^k = d\mathcal{P}_{ij}^k$ and memory can be saved. If the \inpyth{low_memory_version} is chosen, then we simply use the slower finite element assembly
\vskip 1ex
\noindent
\begin{minipage}{\columnwidth}
\begin{python}
assemble((1/dt)*div(u_)*q*dx)
\end{python}
\end{minipage}
\vskip 1ex

The final step for the fractional step solver is the velocity update that can be written for component $k$ as
\begin{equation}
 M_{ij}\,\mathcal{U}_j^{k,n} =  M_{ij}\,\mathcal{U}_j^{k, I} - \triangle \text{t} \,d\mathcal{P}_{ij}^{k} \mathcal{P}_j^{n-1/2}, 
 \label{eq:velocity_update}
\end{equation}
where $\mathcal{U}_j^{k, I}$ and $\mathcal{P}_j^{n-1/2}$ now are the known degrees of freedom of tentative velocity and pressure respectively, whereas $\mathcal{U}_j^{k, n}$ represent the unknowns. The velocity update requires a linear algebra Krylov or direct solve and as such it is quite expensive even though the equation is cheap to assemble. For this reason the velocity update has an additional option to use either a weighted gradient matrix\footnote{Requires the \emph{fenicstools} \cite{fenicstools} package.} $\mathcal{G}_{ij}^k$ or lumping of the mass matrix, that allows the update to be performed directly
\begin{equation}
\mathcal{U}_i^{k, n} =  \mathcal{U}_i^{k, I} - \triangle \text{t} \, \mathcal{G}_{ij}^k \, \mathcal{P}_j^{n-1/2}, \,\text{for}\, i = 1, \ldots, N_u.
\end{equation}
The parameter used to enable the direct approach is \inpyth{NS_parameters["velocity_update_type"]} that can be set to \inpyth{"gradient_matrix"} or \inpyth{"lumping"}.

\section{Verification of implementation}
\label{sec:benchmark}

The fractional step algorithm implemented in \inpyth{NSfracStep} is targeting transient flows in large-scale applications, with turbulent as well as laminar or transitional flow. It is not intended to be used as a steady state solver.\footnote{As of 14 Sep 2014 Oasis ships with a coupled steady state solver for this purpose.}  \emph{Oasis} has previously been used to study, e.g.,  blood flow in highly complex intracranial aneurysms \cite{Steinman-2013, kvs-2014}, where the results compare very well with, e.g., the spectral element code NEKTAR \cite{nektar}. Simulations by Steinman and Valen-Sendstad \cite{kvs-2014} are also commented by Ventikos \cite{ventikos-14}, who state this is "the right way to do it" - referring to the need for highly resolved CFD simulations of transitional blood flow in aneurysms. 

Considering the end use of the solver in  biomedical applications and research, it is essential that we establish the accuracy as well as the efficiency of the solver.

\subsection{2D Taylor Green flow}
\label{sec:2DTG}
Two dimensional Taylor-Green flow is one of very few non-trivial analytical and transient solutions to the Navier-Stokes equations. For this reason it is often used for verification of computer codes. The implementation can be found in \inpyth{Oasis/problems/TaylorGreen.py} and the Taylor Green solution reads
\begin{align}
  \bm{u}_e &= \Big(- \sin (\pi y) \, \cos (\pi x)\, \exp(-2 \pi^2 \nu t), \\
    &\quad\quad\quad \sin (\pi x) \, \cos (\pi y)\, \exp(-2 \pi^2 \nu t) \Big), \\
    p_e &= - \frac{1}{4}\left(\cos (2\pi x)+\cos(2 \pi y) \right) \exp(-4 \pi^2 \nu t),
\end{align}
on the doubly periodic domain $(x, y) = [0, 2]\times [0, 2]$. The analytical solution is used to initialize the solver and to compute the norms of the error, i.e., $||\bm{u}-\bm{u}_e||_h$ and $\| p - p_e\|_h$, where $\| \cdot \|_h$ represents an L2 error norm. The mesh consists entirely of right triangles and is uniform in both spatial directions. The mesh size $h$ is computed as two times the circumradium of a triangle. The kinematic viscosity is set to $\nu=0.01$ and time is integrated for $t=[0, 1]$ with a short timestep ($\triangle \text{t} = 0.001$) to practically eliminate temporal integration errors. The solver is run for a range of mesh sizes and the order of convergence is shown in Table~\ref{tab:orderP2P21}. The velocity is either piecewise quadratic (P2) or piecewise linear (P1), whereas the pressure is always piecewise linear. The P2P1 solver achieves fourth order accuracy in velocity and second order in pressure, whereas the P1P1 solver is second order accurate in both. Note that the fourth order in velocity is due to superconvergence \cite{raey} and it will drop to three for a mesh that is not regularly sized and aligned with the coordinate axis. The order of the L2 error ($k$) is computed by comparing the error norm of two consecutive discretization levels $i$ and $i-1$, and assuming that the error can be written as $E_i = Ch_i^k$, where $C$ is an arbitrary constant. Comparing $E_i=Ch_i^k$ and $E_{i-1}=Ch_{i-1}^k$ we can isolate $k = \ln(E_i/E_{i-1})/\ln(h_i/h_{i-1})$.
\begin{table}[t!]
\caption{Taylor-Green flow convergence errors $O(h^k)$, where $h$ and $k$ are mesh size and order of convergence respectively.  $\| \cdot \|_h$ represents an L2 norm. The velocity is either quadratic (P2) or linear (P1), whereas the pressure is always linear (P1).}
\label{tab:orderP2P21}
\begin{tabular}{p{4em} p{4em} p{2em} p{4em} p{2em}}
    \\
   P2P1 \\
\rule{0pt}{3ex}    
   h & $\| \bm{u} -\bm{u}_e\|_h $ & k & $\| p-p_e \|_h $  & k  \\
   \hline \\   
   2.83E-01 &   2.14E-02 & -  &   1.81E-02 & - \\ 
   1.41E-01 &   1.44E-03 & 3.89  &   5.49E-03 & 1.72 \\ 
   9.43E-02 &   2.84E-04 & 4.01  &   2.46E-03 & 1.97 \\ 
   7.07E-02 &   8.94E-05 & 4.01  &   1.39E-03 & 2.00 \\ 
   5.66E-02 &   3.65E-05 & 4.01  &   8.88E-04 & 2.00 \\ 
  \hline \\
  P1P1 \\
\rule{0pt}{3ex}    
   h & $\| \bm{u} -\bm{u}_e\|_h $ & k & $\| p-p_e \|_h $  & k  \\
   \hline \\
     2.83E-01 &   9.31E-03 & -  &   4.97E-03 & - \\ 
   1.41E-01 &   2.36E-03 & 1.98  &   1.55E-03 & 1.68 \\ 
   9.43E-02 &   1.06E-03 & 1.98  &   7.12E-04 & 1.92 \\ 
   7.07E-02 &   5.98E-04 & 1.99  &   4.05E-04 & 1.97 \\ 
   5.66E-02 &   3.83E-04 & 1.99  &   2.60E-04 & 1.98 \\ 
   \hline
\end{tabular}
\end{table}

To verify the convergence of the transient fractional step scheme, we isolate temporal errors by practically eliminating spatial discretization errors through the use of high order P4 and P3 elements for velocity and pressure respectively. The solver is then run for a range of time step sizes for $t=[0, 6]$. The error norms at the end of the runs are shown in Table~\ref{tab:orderP4P3} indicating that both pressure and velocity achieve second order accuracy in time.  Note that in Table \ref{tab:orderP4P3}, the order of the error is computed from $E_i=Cdt_i^k$, where $dt_i$ is the time step used at level $i$.
\begin{table}[t!]
\caption{Taylor-Green flow convergence errors $O(dt^k)$, where $dt$ and $k$ are time step and order of convergence respectively. The velocity uses Lagrange elements of degree four (P4), whereas the pressure uses third degree (P3).}
\label{tab:orderP4P3}
\begin{tabular}{p{4em} p{4em} p{2em} p{4em} p{2em}}
    \\
   P4P3 \\
\rule{0pt}{3ex}    
   dt & ${\| \bm{u} -\bm{u}_e\|_h} $ & k & ${\| p-p_e \|_h} $  & k  \\
   \hline \\   
   5.00E-01 &   5.08E-01 & -  &   1.29E+00 & - \\ 
   2.50E-01 &   1.36E-01 & 1.91  &   2.97E-01 & 2.11 \\ 
   1.25E-01 &   3.42E-02 & 1.99  &   7.12E-02 & 2.06 \\ 
   6.25E-02 &   8.62E-03 & 1.99  &   1.77E-02 & 2.01 \\ 
   3.12E-02 &   2.17E-03 & 1.99  &   4.41E-03 & 2.00 \\ 
  \hline \\
\end{tabular}
\end{table}

\subsection{Turbulent channel flow}
\label{sec:channel}
\begin{figure}[t!]
\includegraphics[scale=0.5]{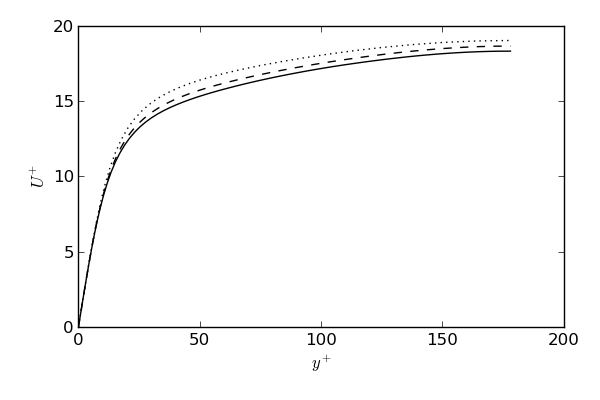}
\caption{Mean velocity in $x$-direction normalized by $u_{\tau}$ as a function of scaled distance to the wall $y^+$. Dotted and dashed curves are computed with \emph{Oasis} using respectively $6\cdot 64^3$ and $6\cdot 128^3$ computational cells. The black curve is the reference solution from MKM.}
\label{fig:Umean}
\end{figure}

The second test case is a direct numerical simulation\footnote{A direct numerical simulation indicates a simulation where all scales of turbulence have been resolved.} of turbulent, fully developed, plane channel flow. The flow is bounded between two parallel planes located at $y = \pm 1$ and is periodic in the $x$ and $z$ directions. The flow is driven by an applied constant pressure gradient (forcing) in the $x$-direction. This flow has been studied extensively with numerous CFD-codes, often using spectral accuracy since it is of primary importance to capture the rate of dissipation of turbulent kinetic energy, allowing no (or very little) numerical diffusion. To verify our implementation we will here attempt to reproduce the classical simulations of Moser, Kim and Mansour (MKM, \cite{mkm99}) for $Re_{\tau} = 180$, based on the wall friction velocity $u_{\tau}=\sqrt{\nu \partial u / \partial y}_{wall}$. The computational box is of size $L_x=4 \pi, L_y = 2$ and $ L_z=4 \pi / 3$. The resolution of MKM was a box of size $128^3$, uniform in $x$ and $z$-directions and skewed towards the walls using Chebyshev points in the $y$-direction. In this test we use one under-resolved box of size $64^3$ and one of the same size as MKM to show convergence towards the correct solution. Since each hexahedron is further divided into 6 tetrahedrons, this corresponds to $6\cdot64^3$ and $6\cdot128^3$ finite elements\footnote{Due to two periodic directions the number of degrees of freedom for the fine mesh are $128\cdot 129 \cdot 128$ for each velocity component and pressure, which is the same as used by MKM.}. MKM performed their simulations using spectral accuracy with Fourier representation in the periodic directions and a Chebyshev-tau formulation in the $y$-direction. Here we use piecewise linear Lagrange elements (P1P1) of second order accuracy. The creation of the mesh and boundary conditions in module \inpyth{problems/Channel.py} is shown in Fig.~\ref{fig:channel}.
\begin{figure}[t!]
\begin{python}[t!]
from numpy import arctan, pi
N = 128
Lx, Ly, Lz = 2.0*pi, 1.0, 2.0*pi/3.0
def mesh(Nx, Ny, Nz, **params):
    m = BoxMesh(0, -Ly, -Lz, Lx, Ly, Lz, 
                N, N, N)
    x = m.coordinates()
    x[:, 1] = arctan(pi*(x[:, 1])) / arctan(pi) 

nu = 2.e-5
Re_tau = 395.
NS_parameters.update(
  nu = nu,
  Re_tau = Re_tau,
  dt = 0.05,
  velocity_degree = 1,
  folder = "channel_results",
  use_krylov_solvers = True)

def walls(x, on_boundary):
    return (on_boundary and 
            near((x[1]+Ly)*(x[1]-Ly), 0.0))

def create_bcs(V, u_components, **NS_name):
    bc = {ui: [DirichletBC(V, 0, walls)] 
               for ui in u_components]}
    bcs['p'] = []    
    return bcs

utau = nu * Re_tau
def body_force(**NS_namespace):
    return Constant((utau**2, 0., 0.))
\end{python}
\caption{Implementation of the Channel problem.}
\label{fig:channel}
\end{figure}
The sampling of statistics is performed using routines from the  \emph{fenicstools} \cite{fenicstools} package and are not described in detail here. Reference is given to the complete source code in \inpyth{problems/Channel.py} in the \emph{Oasis} repository. Figure \ref{fig:Umean} shows the statistically converged mean velocity in the $x$-direction across the channel normalized by $u_{\tau}$. The black curve shows the spectral solution of MKM. The dashed and dotted curves show, respectively, the \emph{Oasis} solution using $6\cdot 64^3$ and $6\cdot 128^3$ computational cells. The coarse solution represents an underresolved simulation where the sharpest velocity gradients cannot be captured. The total amount of dissipation within the flow is thus underpredicted and the mean predicted velocity is consequently higher than it should be. This result is in agreement with the understanding of underresolved Large Eddy Simulations (LES) of turbulent flows, that in effect adds viscosity to the large resolved scales to counteract the lack of dissipation from the unresolved small scales. Hence, simply by increasing the kinematic viscosity, the predicted mean flow could be forced closer to the spectral DNS solution seen in Fig.~\ref{fig:Umean}. Another option is, of course, to refine the mesh and thereby resolve the smallest scales. As expected, we see in Fig.~\ref{fig:Umean} that the $6\cdot 128^3$ simulations are in much closer agreement with the spectral DNS. There is still a slight mismatch, though, that should be attributed to the lower order of the \emph{Oasis} solver, incapable of capturing all the finest scales of turbulence. It is worth mentioning that the Galerkin finite element method used by \emph{Oasis} contains no, or very little, numerical diffusion. A dissipative solver, like, e.g., a finite volume using an upwind scheme or a monotonically integrated implicit LES \cite{fureby99}, would have the same effect as a LES model that adds viscosity and as such could lead to coarse simulations with mean velocity profiles closer to MKM.

\begin{figure}
\includegraphics[scale=0.5]{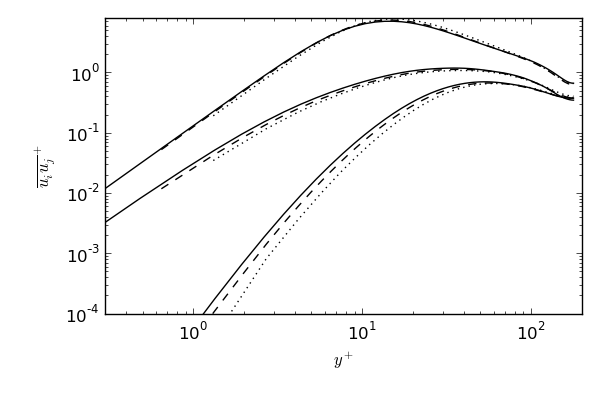}
\caption{Normal Reynolds stresses scaled by $u_{\tau}^2$ shown as functions of scaled distance to the wall $y^+$. Dotted and dashed curves are computed with \emph{Oasis} using respectively $6\cdot 64^3$ and $6\cdot 128^3$ computational cells. The black curves are from the reference solution of MKM. The three different profiles represent, in decreasing magnitude, the normal stresses $\overline{uu}^+, \overline{ww}^+$ and $\overline{vv}^+$, where $u, v$ and $w$ are velocity fluctuations in $x, y$ and $z$ directions respectively.}  
\label{fig:Reynoldsstress}
\end{figure}

Figure \ref{fig:Reynoldsstress} shows the normal, non-dimensionalized, Reynolds stresses. The results confirm that the underresolved stresses are underpredicted close to the wall, whereas the fine simulations converge towards the spectral MKM results.

The channel simulations do not require more computational power than can be provided by a relatively new laptop computer. However, since these simulations are run for more than 30000 timesteps to sample statistics, we have performed parallel computations on the Abel supercomputer at the University of Oslo. The simulations scale weakly when approximately 200,000 elements are used per CPU, and thus we have run our simulations using 8 CPUs for the coarse mesh and 64 for the fine, which contains 12.5 million tetrahedrons. Simulations for the fine grid take approximately 1.5-1.7 seconds real time per computational timestep depending on traffic and distribution on Abel (20-25 \% lower for the coarse simulations) and thus close to 12 hours for the entire test (30000 timesteps). Approximately 75 \% of the computing time is spent in the linear algebra backend's iterative Krylov solvers and assembling of the coefficient matrix, as detailed in Sec.~\ref{sec:hpc}, is responsible for most of the remaining time. The backend (here PETSc) and the iterative linear algebra solvers are thus key to performance. For the tentative velocity computations we have used a stabilized version of a biconjugate gradient squared solver \cite{bicgstab} with a very cheap (fast, low memory) Jacobi preconditioner imported from method \inpyth{get_solvers}, where it is specified as \inpyth{KrylovSolver('bicgstab', 'jacobi')}. This choice is justified since the tentative velocity coefficient matrix is diagonally dominant due to the short timesteps, and each solve requires approximately 10 iterations to converge (the same number for coarse and fine). The pressure coefficient matrix represents a symmetric and elliptic system and thus we choose a solver based on minimal residuals \cite{minres} and the hypre \cite{hypre} algebraic multigrid preconditioner (\inpyth{KrylovSolver('minres', 'hypre_amg')}). The pressure solver uses an average of 6 iterations (same for coarse and fine) to converge to a given tolerance. The velocity update is computed using a lumped mass matrix and no linear algebra solver is thus required for this final step.

\section{Concluding notes on performance} 
\label{sec:conclusions}

The computational speed of any implicit, large-scale Navier-Stokes solver is determined by many competing factors, but most likely it will be limited by hardware and by routines for setting up (assembly) and solving for its linear algebra subsystems. In \emph{Oasis}, and many comparable Navier-Stokes solvers, the linear algebra is performed through routines provided by a backend (here PETSc) and are thus arguably beyond our control. Accepting that we cannot do better than limits imposed  by hardware and the backend, the best we can really hope for through high-level implementations is to eliminate the cost of assembly. In \emph{Oasis} we take all conceivable measures to do just this, as well as even reducing the number of required linear algebra solves. As mentioned in the previous section, for the turbulent channel case with 12.5 mill. tetrahedrons, 75 \% of the computational time was found spent inside very efficient Krylov solvers and we are thus, arguably, pushing at the very boundaries of what may be achieved by a solver developed with similar numerical schemes, using the same backend. 

To further support this claim, without making a complete comparison in terms of accuracy, we have also set up and tested the channel simulations on Abel for two low-level, second order accurate, semi-implicit, fractional step solvers, OpenFOAM\cite{openfoam} and CDP \cite{CDP}, that both are targeting high performance on massively parallel clusters. We used \emph{channelFoam}, distributed with OpenFOAM version 2.2.1 \cite{channelfoam} and 2.5.0 version of CDP (requires license). The \emph{channelFoam} LES solver was modified slightly to run with constant viscosity, and parameters were set to match the finest channel simulations using $128^3$ hexahedral cells. OpenFOAM used for the tentative velocity a biconjugate gradient \cite{bicgstab} solver with a diagonal incomplete-LU preconditioner. For the pressure a conjugate gradient solver was used with a diagonal incomplete Cholesky preconditioner. The CDP solver was set up with the same hexahedral mesh as OpenFOAM, using no model for the LES subgrid viscosity. The linear solvers used by CDP were very similar to those used by \emph{Oasis}, with a Jacobi preconditioned biconjugate gradient solver for the tentative velocity and a generalized minimum residual method \cite{gmres} with the hypre algebraic multigrid preconditioner.  Depending on traffic on Abel, both CDP and \emph{channelFoam} required approximately 1.4-1.7 seconds real time per timestep, which is very close to the speed obtained by \emph{Oasis}. For both CDP and OpenFOAM speed was strongly dominated by the Krylov solvers and both showed the same type of weak scaling as \emph{Oasis} on the Abel supercomputer. 

%In other words, there is very little overhead in the high-level finite element assembly performed by \emph{Oasis} and further performance can only be achieved by boosting the excellent KSP (Krylov Subspace iterative methods and Preconditioners) package provided by PETSc. 

%To further put the solver's performance into perspective, we have compared \emph{Oasis} to the similarly unstructured and implicit low-level C++ solver \emph{channelFoam}, distributed with OpenFOAM version 2.2.1 \cite{channelfoam}. Like FEniCS, OpenFOAM is an open source toolbox for solving partial differential equations. Unlike FEniCS, which is completely general, OpenFOAM is targeted specifically at CFD. In our performance test we modified the channelFoam LES solver slightly to run with constant viscosity, and parameters were set to match the finest channel simulations using $128^3$ hexahedral cells. A head-to-head comparison then revealed that both \emph{channelFoam} and \emph{Oasis} (after the initial set up) required approximately 1.5 seconds per timestep.


\section*{Acknowledgements}
This work has been supported by a Center of Excellence grant from the Research Council of Norway to the Center for Biomedical Computing at Simula Research Laboratory.



%% References


\bibliographystyle{elsarticle-num}
\bibliography{bib}

\begin{thebibliography}{10}
\expandafter\ifx\csname url\endcsname\relax
  \def\url#1{\texttt{#1}}\fi
\expandafter\ifx\csname urlprefix\endcsname\relax\def\urlprefix{URL }\fi
\expandafter\ifx\csname href\endcsname\relax
  \def\href#1#2{#2} \def\path#1{#1}\fi

\bibitem{vmtk}
\href{http://www.vmtk.org}{{VMTK} - {T}he {V}ascular {M}odeling {T}oolkit}.
\newline\urlprefix\url{http://www.vmtk.org}

\bibitem{gmsh}
\href{http://www.geuz.org/gmsh/}{Gmsh: a three-dimensional finite element mesh
  generator with built-in pre- and post-processing facilities}.
\newline\urlprefix\url{http://www.geuz.org/gmsh/}

\bibitem{cubit}
\href{https://cubit.sandia.gov}{The {CUBIT} geometry and mesh generation
  toolkit}.
\newline\urlprefix\url{https://cubit.sandia.gov}

\bibitem{petsc-web-page}
S.~Balay, et.al, {PETSc} {W}eb page, http://www.mcs.anl.gov/petsc (2013).

\bibitem{trilinos}
M.~A. Heroux, et.al, An overview of the trilinos project, ACM Trans. Math.
  Softw. 31~(3) (2005) 397--423.

\bibitem{openfvm}
\href{http://openfvm.sourceforge.net/}{Open{FVM}}.
\newline\urlprefix\url{http://openfvm.sourceforge.net/}

\bibitem{fluidity}
\href{imperial.ac.uk/earthscienceandengineering}{Fluidity}.
\newline\urlprefix\url{imperial.ac.uk/earthscienceandengineering}

\bibitem{oofem}
\href{http://www.oofem.org/en/oofem.html}{Oofem - object oriented finite
  element solver}.
\newline\urlprefix\url{http://www.oofem.org/en/oofem.html}

\bibitem{fenics}
\href{http://fenicsproject.org}{{FE}ni{CS}}.
\newline\urlprefix\url{http://fenicsproject.org}

\bibitem{alnes14}
M.~S. Aln{\ae}s, A.~Logg, K.~B. {\O}lgaard, M.~E. Rognes, G.~N. Wells., Unified
  form language: A domain-specific language for weak formulations of partial
  differential equations, ACM Transactions on Mathematical Software 40~(2).

\bibitem{Kirby:2006}
R.~C. Kirby, A.~Logg, \href{http://doi.acm.org/10.1145/1163641.1163644}{A
  compiler for variational forms}, ACM Trans. Math. Softw. 32~(3) (2006)
  417--444.
\newblock \href {http://dx.doi.org/10.1145/1163641.1163644}
  {\path{doi:10.1145/1163641.1163644}}.
\newline\urlprefix\url{http://doi.acm.org/10.1145/1163641.1163644}

\bibitem{fenicstutorial}
\href{http://fenicsproject.org/documentation/ tutorial}{{FE}ni{CS} tutorial}.
\newline\urlprefix\url{http://fenicsproject.org/documentation/ tutorial}

\bibitem{christon02}
M.~A. Christon, P.~M. Gresho, S.~B. Sutton, Computational predictability of
  time-dependent natural convection flows in enclosures (including a benchmark
  solution), Int. J. for Num. Meth. in Fluids.

\bibitem{simo94}
J.~Simo, F.~Armero, Unconditional stability and long-term behavior of transient
  algorithms for the incompressible navier-stokes and euler equations, Computer
  Methods in Applied Mechanics and Engineering 111 (1994) 111--154.

\bibitem{piso}
R.~I. Issa, Solution of the implicitly discretized fluid flow equations by
  operator-splitting, Journal of Computational Physics 62 (1985) 40--65.

\bibitem{fluent}
\href{www.ansys.com}{Ansys - fluent}.
\newline\urlprefix\url{www.ansys.com}

\bibitem{starcd}
\href{http://www.cd-adapco.com/products/star-cd}{Star-{CD}}.
\newline\urlprefix\url{http://www.cd-adapco.com/products/star-cd}

\bibitem{openfoam}
\href{www.openfoam.com}{Open{FOAM} - {T}he open source {CFD} toolbox}.
\newline\urlprefix\url{www.openfoam.com}

\bibitem{oasismanual}
\href{https://github.com/mikaem/Oasis/blob/master/ doc/usermanual.pdf}{Oasis
  user manual}.
\newline\urlprefix\url{https://github.com/mikaem/Oasis/blob/master/
  doc/usermanual.pdf}

\bibitem{fenicstools}
\href{https://github.com/mikaem/fenicstools}{fenicstools}.
\newline\urlprefix\url{https://github.com/mikaem/fenicstools}

\bibitem{Steinman-2013}
D.~A. Steinman, et.al, Variability of computational fluid dynamics solutions
  for pressure and flow in a giant aneurysm: The asme 2012 summer
  bioengineering conference cfd challenge, Journal of Biomechanical Engineering
  135~(2).

\bibitem{kvs-2014}
K.~Valen-Sendstad, D.~A. Steinman, Mind the gap: Impact of computational fluid
  dynamics solution strategy on prediction of intracranial aneurysm
  hemodynamics and rupture status indicators, American Journal of
  Neuroradiology 35~(3) (2014) 536--543.

\bibitem{nektar}
\href{http://wwwf.imperial.ac.uk/ssherw/spectralhp/ nektar}{{NEKTAR}}.
\newline\urlprefix\url{http://wwwf.imperial.ac.uk/ssherw/spectralhp/ nektar}

\bibitem{ventikos-14}
Y.~Ventikos, Resolving the issue of resolution (2014).

\bibitem{raey}
F.~Guill{\'e}n-gonz{\'a}lez, G.~Tierra, Superconvergence in velocity and
  pressure for the 3{D} time-dependent {N}avier-{S}tokes {E}quations, SeMA
  Journal 57~(1) (2012) 49--67.

\bibitem{mkm99}
R.~D. Moser, J.~Kim, N.~N. Mansour, Direct numerical simulation of turbulent
  channel flow up to re\_tau = 590, Phys. Fluids 11 (1999) 943--945.

\bibitem{fureby99}
C.~Fureby, F.~F. Grinstein., Monotonically integrated large eddy simulation of
  free shear flows, AIAA Journal 37~(5) (1999) 544--556.

\bibitem{bicgstab}
H.~van~der Vorst, Bi-{CGSTAB}: {A} {F}ast and {S}moothly {C}onverging {V}ariant
  of {B}i-{CG} for the {S}olution of {N}onsymmetric {L}inear {S}ystems, SIAM
  Journal on Scientific and Statistical Computing 13~(2) (1992) 631--644.

\bibitem{minres}
C.~C. Paige, M.~A. Saunders, Solution of sparse indefinite systems of linear
  equations, SIAM J. Numerical Analysis 12 (1975) 617--629.

\bibitem{hypre}
\href{http://acts.nersc.gov/hypre/}{Hypre}.
\newline\urlprefix\url{http://acts.nersc.gov/hypre/}

\bibitem{CDP}
\href{http://web.stanford.edu/group/cits/research/ combustor/cdp.html}{{CDP}}.
\newline\urlprefix\url{http://web.stanford.edu/group/cits/research/
  combustor/cdp.html}

\bibitem{channelfoam}
\href{github.com/OpenFOAM/OpenFOAM-2.1.x/tree/master
  /applications/solvers/incompressible/channelFoam}{github.com/openfoam/openfoam-2.1.x/tree/master
  /applications/solvers/incompressible/channelfoam}.
\newline\urlprefix\url{github.com/OpenFOAM/OpenFOAM-2.1.x/tree/master
  /applications/solvers/incompressible/channelFoam}

\bibitem{gmres}
Y.~Saad, M.~Schultz, Gmres: A generalized minimal residual algorithm for
  solving nonsymmetric linear systems, SIAM Journal on Scientific and
  Statistical Computing 7~(3) (1986) 856--869.

\end{thebibliography}






% \end{thebibliography}

\end{document}